\documentclass[aps,twocolumn,preprintnumbers,superscriptaddress,amsmath,amssymb,amsfonts]{revtex4-2}
\usepackage{graphicx}  
\usepackage{color}
\usepackage[colorlinks,bookmarks=true,citecolor=blue,linkcolor=blue,urlcolor=blue, breaklinks=true]{hyperref}
\usepackage{bm}
\usepackage{amsmath,amssymb}
\begin{document}

\title{Emergent spin-gapped magnetization plateaus  in a spin-1/2 perfect kagome antiferromagnet}

\author{S. Suetsugu}
\thanks{These authors contributed equally to this work.}
\affiliation{Department of Physics, Kyoto University, Kyoto 606-8502, Japan}
\author{T. Asaba}
\thanks{These authors contributed equally to this work.}
\affiliation{Department of Physics, Kyoto University, Kyoto 606-8502, Japan}
\author{Y. Kasahara}
\affiliation{Department of Physics, Kyoto University, Kyoto 606-8502, Japan}
\author{Y. Kohsaka}
\affiliation{Department of Physics, Kyoto University, Kyoto 606-8502, Japan}
\author{K. Totsuka}
\affiliation{Yukawa Institute for Theoretical Physics, Kyoto University, Kyoto 606-8502, Japan}
\author{B. Li}
\affiliation{Wuhan National High Magnetic Field Center and School of Physics, Huazhong University of Science and Technology, 430074 Wuhan, China}
\author{Y. Zhao}
\affiliation{Wuhan National High Magnetic Field Center and School of Physics, Huazhong University of Science and Technology, 430074 Wuhan, China}
\author{Y. Li}
\affiliation{Wuhan National High Magnetic Field Center and School of Physics, Huazhong University of Science and Technology, 430074 Wuhan, China}
\author{M. Tokunaga}
\affiliation{Institute for Solid State Physics, The University of Tokyo, Kashiwa, Chiba 277-8581, Japan}
\author{Y. Matsuda}
\affiliation{Department of Physics, Kyoto University, Kyoto 606-8502, Japan}

\date{\today}



\begin{abstract}
The two-dimensional (2D) spin-1/2 kagome Heisenberg antiferromagnet is believed to host quantum spin liquid (QSL) states with no magnetic order, but its ground state remains largely elusive. An important outstanding question concerns the presence or absence of the 1/9 magnetization plateau, where exotic quantum states, including topological ones, are expected to emerge. Here we report the magnetization of a recently discovered kagome QSL candidate YCu$_3$(OH)$_{6.5}$Br$_{2.5}$ up to 57\,T. Above 50\,T, a clear magnetization plateau at 1/3 of the saturation moment of Cu$^{2+}$ ions is observed, supporting that this material provides an ideal platform for the kagome Heisenberg antiferromagnet. Remarkably, we found another magnetization plateau around 20\,T, which is attributed to the 1/9 plateau. The temperature dependence of this plateau reveals the distinct spin gap, whose magnitude estimated by the plateau width is approximately 10\% of the exchange interaction. The observation of 1/9 and 1/3 plateaus highlights the emergence of novel states in quantum spin systems.
\end{abstract}

\maketitle

Frustrated spin systems offer a rich platform for exotic quantum many-body states, originating from competing interactions and quantum fluctuations. Among such systems, QSLs \cite{PhysRevB.45.12377,balents2010spin} are the most fascinating states, which are one of the most entangled quantum states conceived to date. It is widely believed that long-range quantum entanglement leads to many amazing emergent phenomena, such as fractionalized excitations and topological orders. The spin-1/2 Heisenberg antiferromagnet (AFM) on a 2D kagome lattice serves as a prime example for the QSL. However, despite decades of tremendous research, understanding the nature of the kagome AFM has proved to be one of the most vexing issues in the quantum spin systems.

There are several crucially important but unresolved issues for elucidating the ground state properties of the 2D kagome system. Among them, whether the ground state in zero field is gapped or gapless has been highly controversial \cite{yan2011spin,PhysRevX.7.031020,PhysRevB.100.155142}. The ground state in an external magnetic field has also been largely elusive both theoretically and experimentally. Of particular interest is the magnetization plateaus arising from a field-induced spin gap \cite{nishimoto2013controlling,PhysRevB.93.060407,chen2018thermodynamics,PhysRevB.107.L220401,okuma2019series,PhysRevB.88.144416,PhysRevB.98.094423}. Specifically, it has been discussed theoretically that in an external magnetic field, the kagome AFM may exhibit a series of spin-gapped phases with magnetization plateaus at 1/9, 1/3, 5/9, and 7/9 of the saturation moment. 

Theoretically, some or all of these plateaus are expected to appear as a result of quantum entanglement, rather than simple energetics of classical spins. The 1/3 plateau in the kagome AFM, which is found rather robustly regardless of the theoretical methods used \cite{nishimoto2013controlling,PhysRevB.93.060407,chen2018thermodynamics,PhysRevB.107.L220401,okuma2019series,PhysRevB.88.144416,PhysRevB.98.094423}, has been suggested to appear purely due to quantum mechanical effects, unlike the one of classical origin known for the triangular AFM \cite{chubukov1991quantum}. In stark contrast to the 1/3 plateau, as for the 1/9 plateau, even its existence is a nontrivial problem. The 1/9 plateau state, if exists, is expected to be a highly unusual quantum state including a QSL with a topological order \cite{nishimoto2013controlling}. 

Experimental verification of whether such magnetization plateaus really exist should be the key to exploring the enigmatic ground state phases of the spin-1/2 kagome AFM. However, its elucidation remains a significant challenge because there are no ideal candidate materials for such a system with a QSL ground state. Thus, we may safely conclude that even the existence of the plateaus is still open.

Until now, there are only a few candidate materials for the spin-1/2 kagome AFM. Among them, herbertsmithite ZnCu$_3$(OH)$_6$Cl$_2$ \cite{mendels2010quantum,RevModPhys.88.041002} has been most extensively studied as a canonical candidate for bearing a QSL ground state, because it possesses a perfect kagome structure and does not magnetically order down to 50\,mK \cite{PhysRevLett.98.077204,PhysRevLett.98.107204}. The spinon continuum has also been reported by neutron scattering experiments \cite{han2012fractionalized}, supporting the spin fractionalization. However, the observation of the magnetization plateaus in herbertsmithite has been prevented by multiple factors. Firstly, the intrinsic magnetic properties are significantly influenced by orphan spins \cite{PhysRevLett.100.087202,PhysRevLett.100.077203,nilsen2013low,PhysRevB.94.060409,fu2015evidence,khuntia2020gapless,wang2021emergence} introduced by inevitable replacements of Cu$^{2+}$ and Zn$^{2+}$ ions in and between the 2D kagome layers \cite{shores2005structurally,freedman2010site}. In addition, relatively large exchange interaction in the 2D kagome layers makes it difficult to access the field required to observe the plateaus. In herbertsmithite, an anomaly near 1/3 magnetization has been claimed near 150\,T \cite{PhysRevB.102.104429}, but drawing a definite conclusion is difficult owing to the limitation of experimental resolution. The presence of the 1/3 magnetization plateau has been reported for other kagome AFMs such as volborthite Cu$_3$V$_2$O$_7$(OH)$_2\cdot$2H$_2$O \cite{PhysRevLett.114.227202}, Cd-kapellasite CdCu$_3$(OH)$_6$(NO$_3$)$_2\cdot$H$_2$O \cite{okuma2019series}, and Cs$_2A$Ti$_3$F$_{12}$ ($A=$Li, Na, K) \cite{PhysRevB.94.104432,PhysRevB.100.174401}, but the distortion of a kagome lattice or additional magnetic interactions causing a long-range order may mask the intrinsic nature of the ideal kagome AFM. No signature of the 1/9 plateau has been observed not only in kagome AFMs but also in any insulating spin-1/2 quantum AFMs.

\begin{figure}
	\includegraphics[clip,width=8.5cm]{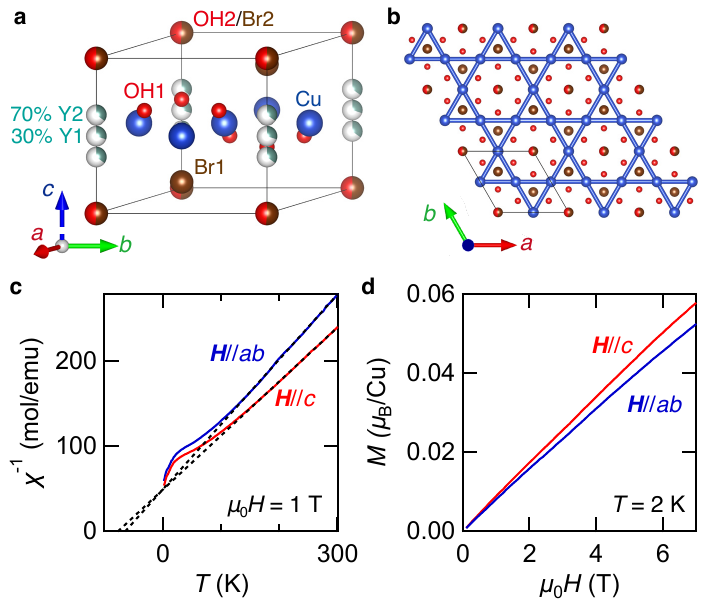}
	\caption{\textbf{Crystal structure and magnetic properties of YCOB.} \textbf{a}, Crystal structure of YCOB. The intersite mixing of OH2 and Br2 displaces 70\% of Y$^{3+}$ ions from the optimal Y1 position on the kagome plane. \textbf{b}, The kagome plane of Cu$^{2+}$ in YCOB. YCOB consists of a 2D perfect kagome lattice of Cu$^{2+}$ ions without antisite mixing by other nonmagnetic ions. \textbf{c}, Temperature dependence of the inverse of the magnetic susceptibility $\chi^{-1}$ at 1\,T for both $\bm{H}||c$ and $\bm{H}||ab$. Linear fits of the high temperature data are denoted by dashed lines. The large negative Curie-Weiss temperatures $\theta_c =-77$\,K and $\theta_{ab}=-65$\,K indicate a predominant antiferromagnetic interaction. \textbf{d}, Field dependence of magnetization $M_{ab}$ and $M_{c}$ at 2\,K up to 7\,T. The small anisotropy $M_{c} / M_{ab} = 1.1$ is observed.
	\label{fig:crystal}
	}
\end{figure}

Recently, a promising candidate of spin-1/2 kagome AFM YCu$_3$(OH)$_{6.5}$Br$_{2.5}$ (YCOB) \cite{chen2020quantum,PhysRevB.105.024418,PhysRevB.105.L121109,PhysRevB.106.L220406,lu2022observation} has been discovered. YCOB consists of a 2D perfect kagome layer of Cu$^{2+}$ ions (Figs.~\ref{fig:crystal}a and b). Despite large exchange interaction $J$ of 60-80\,K, no magnetic order has been observed down to 50\,mK \cite{PhysRevB.105.L121109}. The intersite mixing of OH$^{-}$ and Br$^{-}$ pushes 70\% of Y$^{3+}$ ions away from their ideal position, introducing bond randomness in the exchange interaction \cite{PhysRevB.105.024418}. However, the antisite disorder between magnetic Cu$^{2+}$ and nonmagnetic Y$^{3+}$ is absent due to the very different ionic radii, retaining the perfect kagome lattice of Cu$^{2+}$ intact and free from orphan spins.

In this work, we investigate the magnetization process of YCOB up to 57\,T. We observe a pronounced magnetization plateau at 1/3 of the Cu$^{2+}$ saturation moment, supporting that YCOB is close to the ideal kagome Heisenberg AFM. The most prominent feature is a distinct plateau at 1/9 of the saturation moment. These demonstrate the emergence of unique quantum states in this kagome AFM.

Figure~\ref{fig:crystal}c depicts the temperature ($T$) dependence of the inverse of the magnetic susceptibility $\chi_{c}^{-1}$ and $\chi_{ab}^{-1}$ at $\mu_0H$=1\,T for $\bm{H}||c$ and $\bm{H}||ab$, respectively. At high temperatures, both  $\chi_{c}^{-1}$ and $\chi_{ab}^{-1}$ increase linearly with $T$. Linear fits of the high temperature data  (dashed lines) give the Curie-Weiss temperatures $\theta_c=-77$\,K and $\theta_{ab}=-65$\,K, indicating a predominant antiferromagnetic interaction. As shown in Fig.~\ref{fig:crystal}d, the field dependence of magnetization $M$ exhibits a small anisotropy $M_{c} / M_{ab} = 1.1$.

\begin{figure}
	\includegraphics[clip,width=8cm]{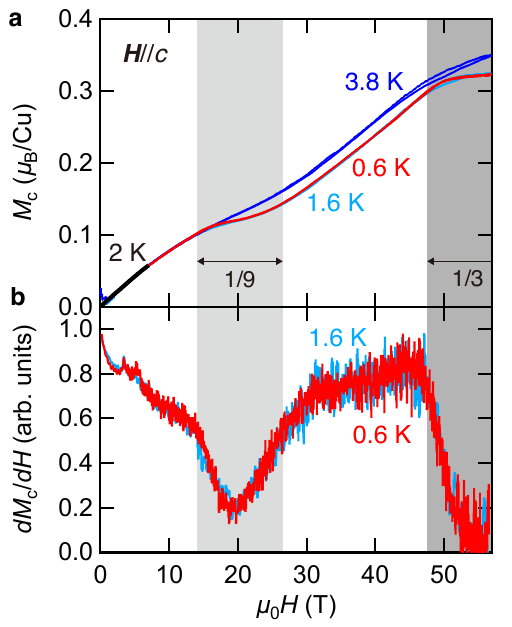}
	\caption{\textbf{Magnetization process of YCOB for $\bm{H}||c$ up to 57\,T.}
	\textbf{a}, Magnetization process at 0.6\,K, 1.6\,K, and 3.8\,K. At 0.6\,K, distinct magnetization plateaus at 1/9 and 1/3 of the Cu$^{2+}$ saturation moment are observed, as highlighted by light and dark gray regions, respectively. The magnetization curve at 0.6\,K nearly perfectly overlaps with that at 1.6\,K in both plateau regions, which provides evidence for the formation of the spin gap. The absolute values were calibrated by the MPMS data at 2\,K (black line). \textbf{b}, The field dependence of the derivative $dM_c/dH$ at 0.6\,K and 1.6\,K. $dM_c/dH$ is significantly reduced within the 1/9 and 1/3 plateaus regions.
	\label{fig:Mc}
	}
\end{figure}

Figure~\ref{fig:Mc}a displays the field ($H$) dependence of $M_{c}$ at several temperatures. The field derivative of the data $dM_{c}/dH$ is also shown in Fig.~\ref{fig:Mc}b. We first discuss the high field feature above 50\,T highlighted by a dark gray region. At the lowest temperature of 0.6\,K, $M_{c}$ flattens out with $H$ above 50\,T, indicating the appearance of a magnetization plateau. The plateau is also confirmed by the sharp reduction of $dM_{c}/dH$, which almost vanishes around 55\,T. Notably,  $M_c$ at the plateau is very close to 1/3 of the fully polarized value (1$\mu_B$/Cu), demonstrating the presence of the 1/3 magnetization plateau. This result provides the first compelling evidence for the 1/3 plateau in the kagome AFM candidate with no magnetic order, revealing that this system serves as an ideal platform for the quest of spin-1/2 kagome Heisenberg AFM.

We now focus on the magnetization behavior below 50\,T. The most salient feature is that the slope of the magnetization flattens out around 20\,T at low temperatures. This is also evident from the strong reduction of $dM_{c}/dH$ in Fig.~\ref{fig:Mc}b. Around 20\,T, $dM_{c}/dH$ shows deep minima, approaching zero. However, unlike to the 1/3 plateau, it does not completely fall to zero, indicating that the magnetization curve has a small but finite slope. We will discuss the possible origins of this later. Remarkably, the absolute value of $M_c$ at 20\,T is close to one-third of that of the 1/3 plateau, \textit{i.e.,} 1/9 of the fully polarized value. We therefore conclude the emergence of 1/9 plateau around 20\,T.

This is further supported by the discussion of the field-induced spin gap for these plateaus. Firstly, the magnetization curve at 1.6\,K nearly perfectly overlaps with that at 0.6\,K for $\bm{H}||c$ in both plateau regions (Figs.~\ref{fig:Mc}a and b). This provides evidence for the formation of the spin gap between the ground state and the first excited state. This strongly supports the emergence of the magnetization plateaus. In addition, the gap size is roughly estimated by the plateau width, since the spin gap is closed when the Zeeman energy $g\mu_\mathrm{B}H$ reaches the gap size. The plateau width around 20\,T is $\sim10$\,T (a light gray region in Figs.~\ref{fig:Mc}a and b), corresponding to the spin gap of $\sim 15$\,K. This value is consistent with the magnetization plateau that becomes less pronounced at 3.8\,K. We also note that this magnitude is roughly consistent with the theoretical value of $\sim 0.1J\approx7$\,K.

\begin{figure}
	\includegraphics[clip,width=8cm]{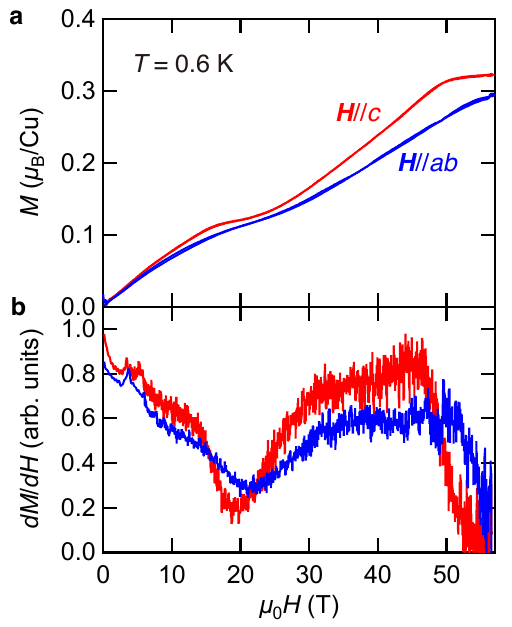}
	\caption{\textbf{Anisotropy of magnetization process in YCOB up to 57\,T.}
	\textbf{a}, Magnetization process at 0.6\,K for $\bm{H}||ab$ and $\bm{H}||c$. Both 1/9 and 1/3 magnetization plateaus are less pronounced for $M_{ab}$ compared to $M_{c}$. \textbf{b}, The field dependence of the derivative $dM/dH$ for the data plotted in (\textbf{a}). For the 1/9 plateau, $dM_{c}/dH$ and $dM_{ab}/dH$ show minima around 19\,T and 21\,T, respectively.
	\label{fig:Mab}
	}
\end{figure}

As shown in Fig.~\ref{fig:Mab}a, both 1/9 and 1/3 magnetization plateaus are observed in $M_{ab}$ around 20\,T and above 50\,T respectively, which is also evident from the reductions of $dM_{ab}/dH$ in Fig.~\ref{fig:Mab}b. The magnetic fields at the center of the 1/9 plateau, determined by the minima of $dM/dH$, are 19\,T and 21\,T for $\bm{H}||c$ and $\bm{H}||ab$, respectively. The nearly 10\% difference, combined with the observed magnetic anisotropy $M_{c}/M_{ab}=1.1$ (see Fig.~\ref{fig:crystal}d), indicates $M_{ab}$ is almost the same value as $M_c$ at the center of the 1/9 plateau. This agreement further supports the presence of the 1/9 plateau, in which the magnitude of the magnetization is independent of the field direction.

The perfect overlap of the magnetization curves at 0.6\,K and 1.6\,K for $\bm{H}||c$ (Figs.~\ref{fig:Mc}a and b) indicates that the finite slope of the magnetization curve at 1/9 plateau is intrinsic, not caused by the thermal broadening. In general, finite slopes of the magnetization plateaus are induced by magnetic interactions that do not conserve the quantum number of spin angular momentum along the field direction. In fact, as discussed above, the system has small but finite magnetic anisotropy and deviates from the isotropic Heisenberg model. In addition, both 1/9 and 1/3 magnetization plateaus are less pronounced in $M_{ab}$ compared to $M_{c}$, as shown in Fig.~\ref{fig:Mab}. A possible origin of this magnetic anisotropy includes Dzyaloshinskii-Moriya interactions or a staggered $g$ tensor, both of which result in non-conservation of the spin angular momentum parallel to the applied field. Moreover, bond randomness inherent to YCOB \cite{PhysRevB.105.024418} may lead to the finite distribution of the spin gap, smearing out the sharp edges of the plateau.

\begin{figure}
	\includegraphics[clip,width=8cm]{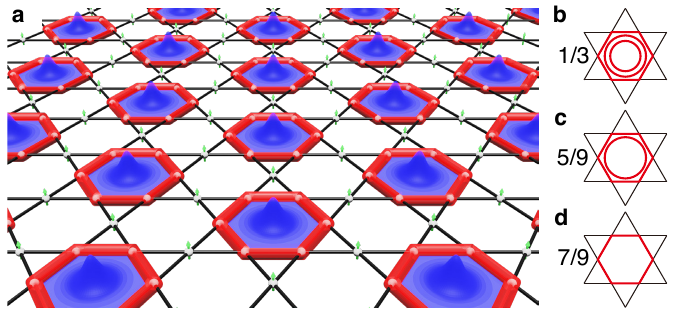}
	\caption{\textbf{Crystal state of localized magnons.}
	\textbf{a} Magnon crystal state at 1/3, 5/9, and 7/9 magnetization plateaus. Six entangled spins (red hexagons) form the extended magnons (blue) within the hexagon of the kagome lattice, that crystallizes into a $\sqrt{3}\times\sqrt{3}$ superstructure. The other spins (green arrows) point in the field direction. \textbf{b}-\textbf{d}, The hexagonal magnons at 1/3, 5/9, and 7/9 plateaus. Three, two, or one magnons (red hexagons and circles) are localized on each hexagon at 1/3, 5/9, and 7/9 plateaus, respectively. 
\label{fig:magnon-crystals}
	}
\end{figure}

The 1/3 plateau in the kagome AFM has been discussed as a quantum mechanical state rather than a classical one. In fact, for the 1/3 plateau, numerical calculations reveal that a crystalline state of entangled magnons living on the individual hexagons gives lower energy than a classical up-up-down spin state \cite{nishimoto2013controlling,chen2018thermodynamics,okuma2019series,PhysRevB.88.144416}. In such a magnon crystal, six spins on each hexagonal plaquette form an entangled state with three resonant magnons, which then crystallize into a $\sqrt{3}\times\sqrt{3}$ superstructure (Figs.~\ref{fig:magnon-crystals}a and b). This picture is robustly inferred from a similar state with one magnon on each hexagonal plaquette (Fig.~\ref{fig:magnon-crystals}d) obtained as the rigorous ground state at the 7/9 plateau \cite{PhysRevLett.88.167207,PhysRevLett.125.117207}, and generalizes to the 5/9 plateau as well (Fig.~\ref{fig:magnon-crystals}c). Therefore, a crystal of emergent highly entangled magnon complexes is anticipated in the observed 1/3 plateau state.

On the other hand,  the 1/9 plateau is hardly accounted for by this magnon crystal picture, suggesting even more exotic states. One intriguing scenario is a QSL state exhibiting a $\mathbb{Z}_3$ topological order, as observed in density matrix renormalization group simulations \cite{nishimoto2013controlling}. Another potential explanation is valence bond crystals suggested by tensor network studies \cite{PhysRevB.93.060407,PhysRevB.107.L220401}, the latter of which concludes an hourglass $\sqrt{3}\times\sqrt{3}$ superstructure with gapless nonmagnetic excitations. Thus, examining the low energy excitations and the spatial symmetry breaking is pivotal in distinguishing these states.

The observation of both 1/3 and 1/9 magnetization plateaus in the spin-1/2 2D quantum magnet with a perfect kagome structure, which is a candidate for QSL, demonstrates that the present system provides an ideal platform for exploring strongly correlated exotic quantum states of matter.

\noindent
{\bf Methods}\\
{\bf Sample growth and characterization}\\
Single crystals of YCOB were grown by a hydrothermal method as reported previously \cite{PhysRevB.105.024418}. Multiple single crystals, aligned along the crystal $c$ axis, were collected for magnetization studies. Magnetization experiments up to 7\,T were conducted by a magnetic property measurement system (MPMS).

\noindent
{\bf High field magnetization experiments}\\
Magnetic fields up to 57\,T were generated using a non-destructive pulse magnet installed at the International MegaGauss Science Laboratory of the Institute for Solid State Physics in the University of Tokyo. Magnetization in pulsed fields was measured by the standard induction method using co-axial pick-up coils. The absolute values of the magnetization were calibrated by the low field data measured by MPMS.

\noindent
{\bf Acknowledgments}\\
We thank A. Miyake and M. G. Yamada for insightful discussions. This work is supported by Grants-in-Aid for Scientific Research (KAKENHI) (Nos. 23K13060 and 23H00089) and on Innovative Areas ``Quantum Liquid Crystals'' (No. JP19H05824) from the Japan Society for the Promotion of Science, and JST CREST (JPMJCR19T5).

\noindent
{\bf Author contributions}\\
B.L, Y.Z., and Y.L prepared and characterized the samples. S.S., T.A., and M.T. performed the magnetization experiments. All authors discussed the results. S.S., T.A., K.T., and Y.M wrote the manuscript with inputs from all authors.

\noindent
{\bf Competing interests}\\
The authors declare no competing financial interests.


\begin{thebibliography}{38}%
	\makeatletter
	\providecommand \@ifxundefined [1]{%
	 \@ifx{#1\undefined}
	}%
	\providecommand \@ifnum [1]{%
	 \ifnum #1\expandafter \@firstoftwo
	 \else \expandafter \@secondoftwo
	 \fi
	}%
	\providecommand \@ifx [1]{%
	 \ifx #1\expandafter \@firstoftwo
	 \else \expandafter \@secondoftwo
	 \fi
	}%
	\providecommand \natexlab [1]{#1}%
	\providecommand \enquote  [1]{``#1''}%
	\providecommand \bibnamefont  [1]{#1}%
	\providecommand \bibfnamefont [1]{#1}%
	\providecommand \citenamefont [1]{#1}%
	\providecommand \href@noop [0]{\@secondoftwo}%
	\providecommand \href [0]{\begingroup \@sanitize@url \@href}%
	\providecommand \@href[1]{\@@startlink{#1}\@@href}%
	\providecommand \@@href[1]{\endgroup#1\@@endlink}%
	\providecommand \@sanitize@url [0]{\catcode `\\12\catcode `\$12\catcode
	  `\&12\catcode `\#12\catcode `\^12\catcode `\_12\catcode `\%12\relax}%
	\providecommand \@@startlink[1]{}%
	\providecommand \@@endlink[0]{}%
	\providecommand \url  [0]{\begingroup\@sanitize@url \@url }%
	\providecommand \@url [1]{\endgroup\@href {#1}{\urlprefix }}%
	\providecommand \urlprefix  [0]{URL }%
	\providecommand \Eprint [0]{\href }%
	\providecommand \doibase [0]{https://doi.org/}%
	\providecommand \selectlanguage [0]{\@gobble}%
	\providecommand \bibinfo  [0]{\@secondoftwo}%
	\providecommand \bibfield  [0]{\@secondoftwo}%
	\providecommand \translation [1]{[#1]}%
	\providecommand \BibitemOpen [0]{}%
	\providecommand \bibitemStop [0]{}%
	\providecommand \bibitemNoStop [0]{.\EOS\space}%
	\providecommand \EOS [0]{\spacefactor3000\relax}%
	\providecommand \BibitemShut  [1]{\csname bibitem#1\endcsname}%
	\let\auto@bib@innerbib\@empty
	\bibitem [{\citenamefont {Sachdev}(1992)}]{PhysRevB.45.12377}%
	  \BibitemOpen
	  \bibfield  {author} {\bibinfo {author} {\bibfnamefont {S.}~\bibnamefont
	  {Sachdev}},\ }\href {https://doi.org/10.1103/PhysRevB.45.12377} {\bibfield
	  {journal} {\bibinfo  {journal} {Phys. Rev. B}\ }\textbf {\bibinfo {volume}
	  {45}},\ \bibinfo {pages} {12377} (\bibinfo {year} {1992})}\BibitemShut
	  {NoStop}%
	\bibitem [{\citenamefont {Balents}(2010)}]{balents2010spin}%
	  \BibitemOpen
	  \bibfield  {author} {\bibinfo {author} {\bibfnamefont {L.}~\bibnamefont
	  {Balents}},\ }\href {https://www.nature.com/articles/nature08917} {\bibfield
	  {journal} {\bibinfo  {journal} {Nature}\ }\textbf {\bibinfo {volume} {464}},\
	  \bibinfo {pages} {199} (\bibinfo {year} {2010})}\BibitemShut {NoStop}%
	\bibitem [{\citenamefont {Yan}\ \emph {et~al.}(2011)\citenamefont {Yan},
	  \citenamefont {Huse},\ and\ \citenamefont {White}}]{yan2011spin}%
	  \BibitemOpen
	  \bibfield  {author} {\bibinfo {author} {\bibfnamefont {S.}~\bibnamefont
	  {Yan}}, \bibinfo {author} {\bibfnamefont {D.~A.}\ \bibnamefont {Huse}},\ and\
	  \bibinfo {author} {\bibfnamefont {S.~R.}\ \bibnamefont {White}},\ }\href
	  {https://www.science.org/doi/10.1126/science.1201080} {\bibfield  {journal}
	  {\bibinfo  {journal} {Science}\ }\textbf {\bibinfo {volume} {332}},\ \bibinfo
	  {pages} {1173} (\bibinfo {year} {2011})}\BibitemShut {NoStop}%
	\bibitem [{\citenamefont {He}\ \emph {et~al.}(2017)\citenamefont {He},
	  \citenamefont {Zaletel}, \citenamefont {Oshikawa},\ and\ \citenamefont
	  {Pollmann}}]{PhysRevX.7.031020}%
	  \BibitemOpen
	  \bibfield  {author} {\bibinfo {author} {\bibfnamefont {Y.-C.}\ \bibnamefont
	  {He}}, \bibinfo {author} {\bibfnamefont {M.~P.}\ \bibnamefont {Zaletel}},
	  \bibinfo {author} {\bibfnamefont {M.}~\bibnamefont {Oshikawa}},\ and\
	  \bibinfo {author} {\bibfnamefont {F.}~\bibnamefont {Pollmann}},\ }\href
	  {https://doi.org/10.1103/PhysRevX.7.031020} {\bibfield  {journal} {\bibinfo
	  {journal} {Phys. Rev. X}\ }\textbf {\bibinfo {volume} {7}},\ \bibinfo {pages}
	  {031020} (\bibinfo {year} {2017})}\BibitemShut {NoStop}%
	\bibitem [{\citenamefont {L\"auchli}\ \emph {et~al.}(2019)\citenamefont
	  {L\"auchli}, \citenamefont {Sudan},\ and\ \citenamefont
	  {Moessner}}]{PhysRevB.100.155142}%
	  \BibitemOpen
	  \bibfield  {author} {\bibinfo {author} {\bibfnamefont {A.~M.}\ \bibnamefont
	  {L\"auchli}}, \bibinfo {author} {\bibfnamefont {J.}~\bibnamefont {Sudan}},\
	  and\ \bibinfo {author} {\bibfnamefont {R.}~\bibnamefont {Moessner}},\ }\href
	  {https://doi.org/10.1103/PhysRevB.100.155142} {\bibfield  {journal} {\bibinfo
	   {journal} {Phys. Rev. B}\ }\textbf {\bibinfo {volume} {100}},\ \bibinfo
	  {pages} {155142} (\bibinfo {year} {2019})}\BibitemShut {NoStop}%
	\bibitem [{\citenamefont {Nishimoto}\ \emph {et~al.}(2013)\citenamefont
	  {Nishimoto}, \citenamefont {Shibata},\ and\ \citenamefont
	  {Hotta}}]{nishimoto2013controlling}%
	  \BibitemOpen
	  \bibfield  {author} {\bibinfo {author} {\bibfnamefont {S.}~\bibnamefont
	  {Nishimoto}}, \bibinfo {author} {\bibfnamefont {N.}~\bibnamefont {Shibata}},\
	  and\ \bibinfo {author} {\bibfnamefont {C.}~\bibnamefont {Hotta}},\ }\href
	  {https://www.nature.com/articles/ncomms3287} {\bibfield  {journal} {\bibinfo
	  {journal} {Nat. Commun.}\ }\textbf {\bibinfo {volume} {4}},\ \bibinfo {pages}
	  {2287} (\bibinfo {year} {2013})}\BibitemShut {NoStop}%
	\bibitem [{\citenamefont {Picot}\ \emph {et~al.}(2016)\citenamefont {Picot},
	  \citenamefont {Ziegler}, \citenamefont {Or\'us},\ and\ \citenamefont
	  {Poilblanc}}]{PhysRevB.93.060407}%
	  \BibitemOpen
	  \bibfield  {author} {\bibinfo {author} {\bibfnamefont {T.}~\bibnamefont
	  {Picot}}, \bibinfo {author} {\bibfnamefont {M.}~\bibnamefont {Ziegler}},
	  \bibinfo {author} {\bibfnamefont {R.}~\bibnamefont {Or\'us}},\ and\ \bibinfo
	  {author} {\bibfnamefont {D.}~\bibnamefont {Poilblanc}},\ }\href
	  {https://doi.org/10.1103/PhysRevB.93.060407} {\bibfield  {journal} {\bibinfo
	  {journal} {Phys. Rev. B}\ }\textbf {\bibinfo {volume} {93}},\ \bibinfo
	  {pages} {060407} (\bibinfo {year} {2016})}\BibitemShut {NoStop}%
	\bibitem [{\citenamefont {Chen}\ \emph {et~al.}(2018)\citenamefont {Chen},
	  \citenamefont {Ran}, \citenamefont {Liu}, \citenamefont {Peng}, \citenamefont
	  {Huang},\ and\ \citenamefont {Su}}]{chen2018thermodynamics}%
	  \BibitemOpen
	  \bibfield  {author} {\bibinfo {author} {\bibfnamefont {X.}~\bibnamefont
	  {Chen}}, \bibinfo {author} {\bibfnamefont {S.-J.}\ \bibnamefont {Ran}},
	  \bibinfo {author} {\bibfnamefont {T.}~\bibnamefont {Liu}}, \bibinfo {author}
	  {\bibfnamefont {C.}~\bibnamefont {Peng}}, \bibinfo {author} {\bibfnamefont
	  {Y.-Z.}\ \bibnamefont {Huang}},\ and\ \bibinfo {author} {\bibfnamefont
	  {G.}~\bibnamefont {Su}},\ }\href
	  {https://www.sciencedirect.com/science/article/abs/pii/S2095927318305267}
	  {\bibfield  {journal} {\bibinfo  {journal} {Sci. Bull.}\ }\textbf {\bibinfo
	  {volume} {63}},\ \bibinfo {pages} {1545} (\bibinfo {year}
	  {2018})}\BibitemShut {NoStop}%
	\bibitem [{\citenamefont {Fang}\ \emph {et~al.}(2023)\citenamefont {Fang},
	  \citenamefont {Xi}, \citenamefont {Ran},\ and\ \citenamefont
	  {Su}}]{PhysRevB.107.L220401}%
	  \BibitemOpen
	  \bibfield  {author} {\bibinfo {author} {\bibfnamefont {D.-z.}\ \bibnamefont
	  {Fang}}, \bibinfo {author} {\bibfnamefont {N.}~\bibnamefont {Xi}}, \bibinfo
	  {author} {\bibfnamefont {S.-J.}\ \bibnamefont {Ran}},\ and\ \bibinfo {author}
	  {\bibfnamefont {G.}~\bibnamefont {Su}},\ }\href
	  {https://doi.org/10.1103/PhysRevB.107.L220401} {\bibfield  {journal}
	  {\bibinfo  {journal} {Phys. Rev. B}\ }\textbf {\bibinfo {volume} {107}},\
	  \bibinfo {pages} {L220401} (\bibinfo {year} {2023})}\BibitemShut {NoStop}%
	\bibitem [{\citenamefont {Okuma}\ \emph {et~al.}(2019)\citenamefont {Okuma},
	  \citenamefont {Nakamura}, \citenamefont {Okubo}, \citenamefont {Miyake},
	  \citenamefont {Matsuo}, \citenamefont {Kindo}, \citenamefont {Tokunaga},
	  \citenamefont {Kawashima}, \citenamefont {Takeyama},\ and\ \citenamefont
	  {Hiroi}}]{okuma2019series}%
	  \BibitemOpen
	  \bibfield  {author} {\bibinfo {author} {\bibfnamefont {R.}~\bibnamefont
	  {Okuma}}, \bibinfo {author} {\bibfnamefont {D.}~\bibnamefont {Nakamura}},
	  \bibinfo {author} {\bibfnamefont {T.}~\bibnamefont {Okubo}}, \bibinfo
	  {author} {\bibfnamefont {A.}~\bibnamefont {Miyake}}, \bibinfo {author}
	  {\bibfnamefont {A.}~\bibnamefont {Matsuo}}, \bibinfo {author} {\bibfnamefont
	  {K.}~\bibnamefont {Kindo}}, \bibinfo {author} {\bibfnamefont
	  {M.}~\bibnamefont {Tokunaga}}, \bibinfo {author} {\bibfnamefont
	  {N.}~\bibnamefont {Kawashima}}, \bibinfo {author} {\bibfnamefont
	  {S.}~\bibnamefont {Takeyama}},\ and\ \bibinfo {author} {\bibfnamefont
	  {Z.}~\bibnamefont {Hiroi}},\ }\href
	  {https://www.nature.com/articles/s41467-019-09063-7} {\bibfield  {journal}
	  {\bibinfo  {journal} {Nat. Commun.}\ }\textbf {\bibinfo {volume} {10}},\
	  \bibinfo {pages} {1229} (\bibinfo {year} {2019})}\BibitemShut {NoStop}%
	\bibitem [{\citenamefont {Capponi}\ \emph {et~al.}(2013)\citenamefont
	  {Capponi}, \citenamefont {Derzhko}, \citenamefont {Honecker}, \citenamefont
	  {L\"auchli},\ and\ \citenamefont {Richter}}]{PhysRevB.88.144416}%
	  \BibitemOpen
	  \bibfield  {author} {\bibinfo {author} {\bibfnamefont {S.}~\bibnamefont
	  {Capponi}}, \bibinfo {author} {\bibfnamefont {O.}~\bibnamefont {Derzhko}},
	  \bibinfo {author} {\bibfnamefont {A.}~\bibnamefont {Honecker}}, \bibinfo
	  {author} {\bibfnamefont {A.~M.}\ \bibnamefont {L\"auchli}},\ and\ \bibinfo
	  {author} {\bibfnamefont {J.}~\bibnamefont {Richter}},\ }\href
	  {https://doi.org/10.1103/PhysRevB.88.144416} {\bibfield  {journal} {\bibinfo
	  {journal} {Phys. Rev. B}\ }\textbf {\bibinfo {volume} {88}},\ \bibinfo
	  {pages} {144416} (\bibinfo {year} {2013})}\BibitemShut {NoStop}%
	\bibitem [{\citenamefont {Schnack}\ \emph {et~al.}(2018)\citenamefont
	  {Schnack}, \citenamefont {Schulenburg},\ and\ \citenamefont
	  {Richter}}]{PhysRevB.98.094423}%
	  \BibitemOpen
	  \bibfield  {author} {\bibinfo {author} {\bibfnamefont {J.}~\bibnamefont
	  {Schnack}}, \bibinfo {author} {\bibfnamefont {J.}~\bibnamefont
	  {Schulenburg}},\ and\ \bibinfo {author} {\bibfnamefont {J.}~\bibnamefont
	  {Richter}},\ }\href {https://doi.org/10.1103/PhysRevB.98.094423} {\bibfield
	  {journal} {\bibinfo  {journal} {Phys. Rev. B}\ }\textbf {\bibinfo {volume}
	  {98}},\ \bibinfo {pages} {094423} (\bibinfo {year} {2018})}\BibitemShut
	  {NoStop}%
	\bibitem [{\citenamefont {Chubukov}\ and\ \citenamefont
	  {Golosov}(1991)}]{chubukov1991quantum}%
	  \BibitemOpen
	  \bibfield  {author} {\bibinfo {author} {\bibfnamefont {A.}~\bibnamefont
	  {Chubukov}}\ and\ \bibinfo {author} {\bibfnamefont {D.}~\bibnamefont
	  {Golosov}},\ }\href
	  {https://iopscience.iop.org/article/10.1088/0953-8984/3/1/005} {\bibfield
	  {journal} {\bibinfo  {journal} {J. Phys. Condens. Matter}\ }\textbf {\bibinfo
	  {volume} {3}},\ \bibinfo {pages} {69} (\bibinfo {year} {1991})}\BibitemShut
	  {NoStop}%
	\bibitem [{\citenamefont {Mendels}\ and\ \citenamefont
	  {Bert}(2010)}]{mendels2010quantum}%
	  \BibitemOpen
	  \bibfield  {author} {\bibinfo {author} {\bibfnamefont {P.}~\bibnamefont
	  {Mendels}}\ and\ \bibinfo {author} {\bibfnamefont {F.}~\bibnamefont {Bert}},\
	  }\href {https://journals.jps.jp/doi/10.1143/JPSJ.79.011001} {\bibfield
	  {journal} {\bibinfo  {journal} {J. Phys. Soc. Jpn.}\ }\textbf {\bibinfo
	  {volume} {79}},\ \bibinfo {pages} {011001} (\bibinfo {year}
	  {2010})}\BibitemShut {NoStop}%
	\bibitem [{\citenamefont {Norman}(2016)}]{RevModPhys.88.041002}%
	  \BibitemOpen
	  \bibfield  {author} {\bibinfo {author} {\bibfnamefont {M.~R.}\ \bibnamefont
	  {Norman}},\ }\href {https://doi.org/10.1103/RevModPhys.88.041002} {\bibfield
	  {journal} {\bibinfo  {journal} {Rev. Mod. Phys.}\ }\textbf {\bibinfo {volume}
	  {88}},\ \bibinfo {pages} {041002} (\bibinfo {year} {2016})}\BibitemShut
	  {NoStop}%
	\bibitem [{\citenamefont {Mendels}\ \emph {et~al.}(2007)\citenamefont
	  {Mendels}, \citenamefont {Bert}, \citenamefont {de~Vries}, \citenamefont
	  {Olariu}, \citenamefont {Harrison}, \citenamefont {Duc}, \citenamefont
	  {Trombe}, \citenamefont {Lord}, \citenamefont {Amato},\ and\ \citenamefont
	  {Baines}}]{PhysRevLett.98.077204}%
	  \BibitemOpen
	  \bibfield  {author} {\bibinfo {author} {\bibfnamefont {P.}~\bibnamefont
	  {Mendels}}, \bibinfo {author} {\bibfnamefont {F.}~\bibnamefont {Bert}},
	  \bibinfo {author} {\bibfnamefont {M.~A.}\ \bibnamefont {de~Vries}}, \bibinfo
	  {author} {\bibfnamefont {A.}~\bibnamefont {Olariu}}, \bibinfo {author}
	  {\bibfnamefont {A.}~\bibnamefont {Harrison}}, \bibinfo {author}
	  {\bibfnamefont {F.}~\bibnamefont {Duc}}, \bibinfo {author} {\bibfnamefont
	  {J.~C.}\ \bibnamefont {Trombe}}, \bibinfo {author} {\bibfnamefont {J.~S.}\
	  \bibnamefont {Lord}}, \bibinfo {author} {\bibfnamefont {A.}~\bibnamefont
	  {Amato}},\ and\ \bibinfo {author} {\bibfnamefont {C.}~\bibnamefont
	  {Baines}},\ }\href {https://doi.org/10.1103/PhysRevLett.98.077204} {\bibfield
	   {journal} {\bibinfo  {journal} {Phys. Rev. Lett.}\ }\textbf {\bibinfo
	  {volume} {98}},\ \bibinfo {pages} {077204} (\bibinfo {year}
	  {2007})}\BibitemShut {NoStop}%
	\bibitem [{\citenamefont {Helton}\ \emph {et~al.}(2007)\citenamefont {Helton},
	  \citenamefont {Matan}, \citenamefont {Shores}, \citenamefont {Nytko},
	  \citenamefont {Bartlett}, \citenamefont {Yoshida}, \citenamefont {Takano},
	  \citenamefont {Suslov}, \citenamefont {Qiu}, \citenamefont {Chung},
	  \citenamefont {Nocera},\ and\ \citenamefont {Lee}}]{PhysRevLett.98.107204}%
	  \BibitemOpen
	  \bibfield  {author} {\bibinfo {author} {\bibfnamefont {J.~S.}\ \bibnamefont
	  {Helton}}, \bibinfo {author} {\bibfnamefont {K.}~\bibnamefont {Matan}},
	  \bibinfo {author} {\bibfnamefont {M.~P.}\ \bibnamefont {Shores}}, \bibinfo
	  {author} {\bibfnamefont {E.~A.}\ \bibnamefont {Nytko}}, \bibinfo {author}
	  {\bibfnamefont {B.~M.}\ \bibnamefont {Bartlett}}, \bibinfo {author}
	  {\bibfnamefont {Y.}~\bibnamefont {Yoshida}}, \bibinfo {author} {\bibfnamefont
	  {Y.}~\bibnamefont {Takano}}, \bibinfo {author} {\bibfnamefont
	  {A.}~\bibnamefont {Suslov}}, \bibinfo {author} {\bibfnamefont
	  {Y.}~\bibnamefont {Qiu}}, \bibinfo {author} {\bibfnamefont {J.-H.}\
	  \bibnamefont {Chung}}, \bibinfo {author} {\bibfnamefont {D.~G.}\ \bibnamefont
	  {Nocera}},\ and\ \bibinfo {author} {\bibfnamefont {Y.~S.}\ \bibnamefont
	  {Lee}},\ }\href {https://doi.org/10.1103/PhysRevLett.98.107204} {\bibfield
	  {journal} {\bibinfo  {journal} {Phys. Rev. Lett.}\ }\textbf {\bibinfo
	  {volume} {98}},\ \bibinfo {pages} {107204} (\bibinfo {year}
	  {2007})}\BibitemShut {NoStop}%
	\bibitem [{\citenamefont {Han}\ \emph {et~al.}(2012)\citenamefont {Han},
	  \citenamefont {Helton}, \citenamefont {Chu}, \citenamefont {Nocera},
	  \citenamefont {Rodriguez-Rivera}, \citenamefont {Broholm},\ and\
	  \citenamefont {Lee}}]{han2012fractionalized}%
	  \BibitemOpen
	  \bibfield  {author} {\bibinfo {author} {\bibfnamefont {T.-H.}\ \bibnamefont
	  {Han}}, \bibinfo {author} {\bibfnamefont {J.~S.}\ \bibnamefont {Helton}},
	  \bibinfo {author} {\bibfnamefont {S.}~\bibnamefont {Chu}}, \bibinfo {author}
	  {\bibfnamefont {D.~G.}\ \bibnamefont {Nocera}}, \bibinfo {author}
	  {\bibfnamefont {J.~A.}\ \bibnamefont {Rodriguez-Rivera}}, \bibinfo {author}
	  {\bibfnamefont {C.}~\bibnamefont {Broholm}},\ and\ \bibinfo {author}
	  {\bibfnamefont {Y.~S.}\ \bibnamefont {Lee}},\ }\href
	  {https://www.nature.com/articles/nature11659} {\bibfield  {journal} {\bibinfo
	   {journal} {Nature}\ }\textbf {\bibinfo {volume} {492}},\ \bibinfo {pages}
	  {406} (\bibinfo {year} {2012})}\BibitemShut {NoStop}%
	\bibitem [{\citenamefont {Olariu}\ \emph {et~al.}(2008)\citenamefont {Olariu},
	  \citenamefont {Mendels}, \citenamefont {Bert}, \citenamefont {Duc},
	  \citenamefont {Trombe}, \citenamefont {de~Vries},\ and\ \citenamefont
	  {Harrison}}]{PhysRevLett.100.087202}%
	  \BibitemOpen
	  \bibfield  {author} {\bibinfo {author} {\bibfnamefont {A.}~\bibnamefont
	  {Olariu}}, \bibinfo {author} {\bibfnamefont {P.}~\bibnamefont {Mendels}},
	  \bibinfo {author} {\bibfnamefont {F.}~\bibnamefont {Bert}}, \bibinfo {author}
	  {\bibfnamefont {F.}~\bibnamefont {Duc}}, \bibinfo {author} {\bibfnamefont
	  {J.~C.}\ \bibnamefont {Trombe}}, \bibinfo {author} {\bibfnamefont {M.~A.}\
	  \bibnamefont {de~Vries}},\ and\ \bibinfo {author} {\bibfnamefont
	  {A.}~\bibnamefont {Harrison}},\ }\href
	  {https://doi.org/10.1103/PhysRevLett.100.087202} {\bibfield  {journal}
	  {\bibinfo  {journal} {Phys. Rev. Lett.}\ }\textbf {\bibinfo {volume} {100}},\
	  \bibinfo {pages} {087202} (\bibinfo {year} {2008})}\BibitemShut {NoStop}%
	\bibitem [{\citenamefont {Imai}\ \emph {et~al.}(2008)\citenamefont {Imai},
	  \citenamefont {Nytko}, \citenamefont {Bartlett}, \citenamefont {Shores},\
	  and\ \citenamefont {Nocera}}]{PhysRevLett.100.077203}%
	  \BibitemOpen
	  \bibfield  {author} {\bibinfo {author} {\bibfnamefont {T.}~\bibnamefont
	  {Imai}}, \bibinfo {author} {\bibfnamefont {E.~A.}\ \bibnamefont {Nytko}},
	  \bibinfo {author} {\bibfnamefont {B.~M.}\ \bibnamefont {Bartlett}}, \bibinfo
	  {author} {\bibfnamefont {M.~P.}\ \bibnamefont {Shores}},\ and\ \bibinfo
	  {author} {\bibfnamefont {D.~G.}\ \bibnamefont {Nocera}},\ }\href
	  {https://doi.org/10.1103/PhysRevLett.100.077203} {\bibfield  {journal}
	  {\bibinfo  {journal} {Phys. Rev. Lett.}\ }\textbf {\bibinfo {volume} {100}},\
	  \bibinfo {pages} {077203} (\bibinfo {year} {2008})}\BibitemShut {NoStop}%
	\bibitem [{\citenamefont {Nilsen}\ \emph {et~al.}(2013)\citenamefont {Nilsen},
	  \citenamefont {De~Vries}, \citenamefont {Stewart}, \citenamefont {Harrison},\
	  and\ \citenamefont {R{\o}nnow}}]{nilsen2013low}%
	  \BibitemOpen
	  \bibfield  {author} {\bibinfo {author} {\bibfnamefont {G.}~\bibnamefont
	  {Nilsen}}, \bibinfo {author} {\bibfnamefont {M.}~\bibnamefont {De~Vries}},
	  \bibinfo {author} {\bibfnamefont {J.~R.}\ \bibnamefont {Stewart}}, \bibinfo
	  {author} {\bibfnamefont {A.}~\bibnamefont {Harrison}},\ and\ \bibinfo
	  {author} {\bibfnamefont {H.}~\bibnamefont {R{\o}nnow}},\ }\href
	  {https://iopscience.iop.org/article/10.1088/0953-8984/25/10/106001}
	  {\bibfield  {journal} {\bibinfo  {journal} {J. Phys. Condens. Matter}\
	  }\textbf {\bibinfo {volume} {25}},\ \bibinfo {pages} {106001} (\bibinfo
	  {year} {2013})}\BibitemShut {NoStop}%
	\bibitem [{\citenamefont {Han}\ \emph {et~al.}(2016)\citenamefont {Han},
	  \citenamefont {Norman}, \citenamefont {Wen}, \citenamefont
	  {Rodriguez-Rivera}, \citenamefont {Helton}, \citenamefont {Broholm},\ and\
	  \citenamefont {Lee}}]{PhysRevB.94.060409}%
	  \BibitemOpen
	  \bibfield  {author} {\bibinfo {author} {\bibfnamefont {T.-H.}\ \bibnamefont
	  {Han}}, \bibinfo {author} {\bibfnamefont {M.~R.}\ \bibnamefont {Norman}},
	  \bibinfo {author} {\bibfnamefont {J.-J.}\ \bibnamefont {Wen}}, \bibinfo
	  {author} {\bibfnamefont {J.~A.}\ \bibnamefont {Rodriguez-Rivera}}, \bibinfo
	  {author} {\bibfnamefont {J.~S.}\ \bibnamefont {Helton}}, \bibinfo {author}
	  {\bibfnamefont {C.}~\bibnamefont {Broholm}},\ and\ \bibinfo {author}
	  {\bibfnamefont {Y.~S.}\ \bibnamefont {Lee}},\ }\href
	  {https://doi.org/10.1103/PhysRevB.94.060409} {\bibfield  {journal} {\bibinfo
	  {journal} {Phys. Rev. B}\ }\textbf {\bibinfo {volume} {94}},\ \bibinfo
	  {pages} {060409} (\bibinfo {year} {2016})}\BibitemShut {NoStop}%
	\bibitem [{\citenamefont {Fu}\ \emph {et~al.}(2015)\citenamefont {Fu},
	  \citenamefont {Imai}, \citenamefont {Han},\ and\ \citenamefont
	  {Lee}}]{fu2015evidence}%
	  \BibitemOpen
	  \bibfield  {author} {\bibinfo {author} {\bibfnamefont {M.}~\bibnamefont
	  {Fu}}, \bibinfo {author} {\bibfnamefont {T.}~\bibnamefont {Imai}}, \bibinfo
	  {author} {\bibfnamefont {T.-H.}\ \bibnamefont {Han}},\ and\ \bibinfo {author}
	  {\bibfnamefont {Y.~S.}\ \bibnamefont {Lee}},\ }\href
	  {https://science.org/doi/10.1126/science.aab2120} {\bibfield  {journal}
	  {\bibinfo  {journal} {Science}\ }\textbf {\bibinfo {volume} {350}},\ \bibinfo
	  {pages} {655} (\bibinfo {year} {2015})}\BibitemShut {NoStop}%
	\bibitem [{\citenamefont {Khuntia}\ \emph {et~al.}(2020)\citenamefont
	  {Khuntia}, \citenamefont {Velazquez}, \citenamefont {Barth{\'e}lemy},
	  \citenamefont {Bert}, \citenamefont {Kermarrec}, \citenamefont {Legros},
	  \citenamefont {Bernu}, \citenamefont {Messio}, \citenamefont {Zorko},\ and\
	  \citenamefont {Mendels}}]{khuntia2020gapless}%
	  \BibitemOpen
	  \bibfield  {author} {\bibinfo {author} {\bibfnamefont {P.}~\bibnamefont
	  {Khuntia}}, \bibinfo {author} {\bibfnamefont {M.}~\bibnamefont {Velazquez}},
	  \bibinfo {author} {\bibfnamefont {Q.}~\bibnamefont {Barth{\'e}lemy}},
	  \bibinfo {author} {\bibfnamefont {F.}~\bibnamefont {Bert}}, \bibinfo {author}
	  {\bibfnamefont {E.}~\bibnamefont {Kermarrec}}, \bibinfo {author}
	  {\bibfnamefont {A.}~\bibnamefont {Legros}}, \bibinfo {author} {\bibfnamefont
	  {B.}~\bibnamefont {Bernu}}, \bibinfo {author} {\bibfnamefont
	  {L.}~\bibnamefont {Messio}}, \bibinfo {author} {\bibfnamefont
	  {A.}~\bibnamefont {Zorko}},\ and\ \bibinfo {author} {\bibfnamefont
	  {P.}~\bibnamefont {Mendels}},\ }\href
	  {https://www.nature.com/articles/s41567-020-0792-1} {\bibfield  {journal}
	  {\bibinfo  {journal} {Nat. Phys.}\ }\textbf {\bibinfo {volume} {16}},\
	  \bibinfo {pages} {469} (\bibinfo {year} {2020})}\BibitemShut {NoStop}%
	\bibitem [{\citenamefont {Wang}\ \emph {et~al.}(2021)\citenamefont {Wang},
	  \citenamefont {Yuan}, \citenamefont {Singer}, \citenamefont {Smaha},
	  \citenamefont {He}, \citenamefont {Wen}, \citenamefont {Lee},\ and\
	  \citenamefont {Imai}}]{wang2021emergence}%
	  \BibitemOpen
	  \bibfield  {author} {\bibinfo {author} {\bibfnamefont {J.}~\bibnamefont
	  {Wang}}, \bibinfo {author} {\bibfnamefont {W.}~\bibnamefont {Yuan}}, \bibinfo
	  {author} {\bibfnamefont {P.~M.}\ \bibnamefont {Singer}}, \bibinfo {author}
	  {\bibfnamefont {R.~W.}\ \bibnamefont {Smaha}}, \bibinfo {author}
	  {\bibfnamefont {W.}~\bibnamefont {He}}, \bibinfo {author} {\bibfnamefont
	  {J.}~\bibnamefont {Wen}}, \bibinfo {author} {\bibfnamefont {Y.~S.}\
	  \bibnamefont {Lee}},\ and\ \bibinfo {author} {\bibfnamefont {T.}~\bibnamefont
	  {Imai}},\ }\href {https://www.nature.com/articles/s41567-021-01310-3}
	  {\bibfield  {journal} {\bibinfo  {journal} {Nat. Phys.}\ }\textbf {\bibinfo
	  {volume} {17}},\ \bibinfo {pages} {1109} (\bibinfo {year}
	  {2021})}\BibitemShut {NoStop}%
	\bibitem [{\citenamefont {Shores}\ \emph {et~al.}(2005)\citenamefont {Shores},
	  \citenamefont {Nytko}, \citenamefont {Bartlett},\ and\ \citenamefont
	  {Nocera}}]{shores2005structurally}%
	  \BibitemOpen
	  \bibfield  {author} {\bibinfo {author} {\bibfnamefont {M.~P.}\ \bibnamefont
	  {Shores}}, \bibinfo {author} {\bibfnamefont {E.~A.}\ \bibnamefont {Nytko}},
	  \bibinfo {author} {\bibfnamefont {B.~M.}\ \bibnamefont {Bartlett}},\ and\
	  \bibinfo {author} {\bibfnamefont {D.~G.}\ \bibnamefont {Nocera}},\ }\href
	  {https://pubs.acs.org/doi/10.1021/ja053891p} {\bibfield  {journal} {\bibinfo
	  {journal} {J. Am. Chem. Soc.}\ }\textbf {\bibinfo {volume} {127}},\ \bibinfo
	  {pages} {13462} (\bibinfo {year} {2005})}\BibitemShut {NoStop}%
	\bibitem [{\citenamefont {Freedman}\ \emph {et~al.}(2010)\citenamefont
	  {Freedman}, \citenamefont {Han}, \citenamefont {Prodi}, \citenamefont
	  {Muller}, \citenamefont {Huang}, \citenamefont {Chen}, \citenamefont {Webb},
	  \citenamefont {Lee}, \citenamefont {McQueen},\ and\ \citenamefont
	  {Nocera}}]{freedman2010site}%
	  \BibitemOpen
	  \bibfield  {author} {\bibinfo {author} {\bibfnamefont {D.~E.}\ \bibnamefont
	  {Freedman}}, \bibinfo {author} {\bibfnamefont {T.~H.}\ \bibnamefont {Han}},
	  \bibinfo {author} {\bibfnamefont {A.}~\bibnamefont {Prodi}}, \bibinfo
	  {author} {\bibfnamefont {P.}~\bibnamefont {Muller}}, \bibinfo {author}
	  {\bibfnamefont {Q.-Z.}\ \bibnamefont {Huang}}, \bibinfo {author}
	  {\bibfnamefont {Y.-S.}\ \bibnamefont {Chen}}, \bibinfo {author}
	  {\bibfnamefont {S.~M.}\ \bibnamefont {Webb}}, \bibinfo {author}
	  {\bibfnamefont {Y.~S.}\ \bibnamefont {Lee}}, \bibinfo {author} {\bibfnamefont
	  {T.~M.}\ \bibnamefont {McQueen}},\ and\ \bibinfo {author} {\bibfnamefont
	  {D.~G.}\ \bibnamefont {Nocera}},\ }\href
	  {https://pubs.acs.org/doi/10.1021/ja1070398} {\bibfield  {journal} {\bibinfo
	  {journal} {J. Am. Chem. Soc.}\ }\textbf {\bibinfo {volume} {132}},\ \bibinfo
	  {pages} {16185} (\bibinfo {year} {2010})}\BibitemShut {NoStop}%
	\bibitem [{\citenamefont {Okuma}\ \emph {et~al.}(2020)\citenamefont {Okuma},
	  \citenamefont {Nakamura},\ and\ \citenamefont
	  {Takeyama}}]{PhysRevB.102.104429}%
	  \BibitemOpen
	  \bibfield  {author} {\bibinfo {author} {\bibfnamefont {R.}~\bibnamefont
	  {Okuma}}, \bibinfo {author} {\bibfnamefont {D.}~\bibnamefont {Nakamura}},\
	  and\ \bibinfo {author} {\bibfnamefont {S.}~\bibnamefont {Takeyama}},\ }\href
	  {https://doi.org/10.1103/PhysRevB.102.104429} {\bibfield  {journal} {\bibinfo
	   {journal} {Phys. Rev. B}\ }\textbf {\bibinfo {volume} {102}},\ \bibinfo
	  {pages} {104429} (\bibinfo {year} {2020})}\BibitemShut {NoStop}%
	\bibitem [{\citenamefont {Ishikawa}\ \emph {et~al.}(2015)\citenamefont
	  {Ishikawa}, \citenamefont {Yoshida}, \citenamefont {Nawa}, \citenamefont
	  {Jeong}, \citenamefont {Kr\"amer}, \citenamefont
	  {Horvati\ifmmode~\acute{c}\else \'{c}\fi{}}, \citenamefont {Berthier},
	  \citenamefont {Takigawa}, \citenamefont {Akaki}, \citenamefont {Miyake},
	  \citenamefont {Tokunaga}, \citenamefont {Kindo}, \citenamefont {Yamaura},
	  \citenamefont {Okamoto},\ and\ \citenamefont
	  {Hiroi}}]{PhysRevLett.114.227202}%
	  \BibitemOpen
	  \bibfield  {author} {\bibinfo {author} {\bibfnamefont {H.}~\bibnamefont
	  {Ishikawa}}, \bibinfo {author} {\bibfnamefont {M.}~\bibnamefont {Yoshida}},
	  \bibinfo {author} {\bibfnamefont {K.}~\bibnamefont {Nawa}}, \bibinfo {author}
	  {\bibfnamefont {M.}~\bibnamefont {Jeong}}, \bibinfo {author} {\bibfnamefont
	  {S.}~\bibnamefont {Kr\"amer}}, \bibinfo {author} {\bibfnamefont
	  {M.}~\bibnamefont {Horvati\ifmmode~\acute{c}\else \'{c}\fi{}}}, \bibinfo
	  {author} {\bibfnamefont {C.}~\bibnamefont {Berthier}}, \bibinfo {author}
	  {\bibfnamefont {M.}~\bibnamefont {Takigawa}}, \bibinfo {author}
	  {\bibfnamefont {M.}~\bibnamefont {Akaki}}, \bibinfo {author} {\bibfnamefont
	  {A.}~\bibnamefont {Miyake}}, \bibinfo {author} {\bibfnamefont
	  {M.}~\bibnamefont {Tokunaga}}, \bibinfo {author} {\bibfnamefont
	  {K.}~\bibnamefont {Kindo}}, \bibinfo {author} {\bibfnamefont
	  {J.}~\bibnamefont {Yamaura}}, \bibinfo {author} {\bibfnamefont
	  {Y.}~\bibnamefont {Okamoto}},\ and\ \bibinfo {author} {\bibfnamefont
	  {Z.}~\bibnamefont {Hiroi}},\ }\href
	  {https://doi.org/10.1103/PhysRevLett.114.227202} {\bibfield  {journal}
	  {\bibinfo  {journal} {Phys. Rev. Lett.}\ }\textbf {\bibinfo {volume} {114}},\
	  \bibinfo {pages} {227202} (\bibinfo {year} {2015})}\BibitemShut {NoStop}%
	\bibitem [{\citenamefont {Goto}\ \emph {et~al.}(2016)\citenamefont {Goto},
	  \citenamefont {Ueda}, \citenamefont {Michioka}, \citenamefont {Matsuo},
	  \citenamefont {Kindo},\ and\ \citenamefont {Yoshimura}}]{PhysRevB.94.104432}%
	  \BibitemOpen
	  \bibfield  {author} {\bibinfo {author} {\bibfnamefont {M.}~\bibnamefont
	  {Goto}}, \bibinfo {author} {\bibfnamefont {H.}~\bibnamefont {Ueda}}, \bibinfo
	  {author} {\bibfnamefont {C.}~\bibnamefont {Michioka}}, \bibinfo {author}
	  {\bibfnamefont {A.}~\bibnamefont {Matsuo}}, \bibinfo {author} {\bibfnamefont
	  {K.}~\bibnamefont {Kindo}},\ and\ \bibinfo {author} {\bibfnamefont
	  {K.}~\bibnamefont {Yoshimura}},\ }\href
	  {https://doi.org/10.1103/PhysRevB.94.104432} {\bibfield  {journal} {\bibinfo
	  {journal} {Phys. Rev. B}\ }\textbf {\bibinfo {volume} {94}},\ \bibinfo
	  {pages} {104432} (\bibinfo {year} {2016})}\BibitemShut {NoStop}%
	\bibitem [{\citenamefont {Shirakami}\ \emph {et~al.}(2019)\citenamefont
	  {Shirakami}, \citenamefont {Ueda}, \citenamefont {Jeschke}, \citenamefont
	  {Nakano}, \citenamefont {Kobayashi}, \citenamefont {Matsuo}, \citenamefont
	  {Sakai}, \citenamefont {Katayama}, \citenamefont {Sawa}, \citenamefont
	  {Kindo}, \citenamefont {Michioka},\ and\ \citenamefont
	  {Yoshimura}}]{PhysRevB.100.174401}%
	  \BibitemOpen
	  \bibfield  {author} {\bibinfo {author} {\bibfnamefont {R.}~\bibnamefont
	  {Shirakami}}, \bibinfo {author} {\bibfnamefont {H.}~\bibnamefont {Ueda}},
	  \bibinfo {author} {\bibfnamefont {H.~O.}\ \bibnamefont {Jeschke}}, \bibinfo
	  {author} {\bibfnamefont {H.}~\bibnamefont {Nakano}}, \bibinfo {author}
	  {\bibfnamefont {S.}~\bibnamefont {Kobayashi}}, \bibinfo {author}
	  {\bibfnamefont {A.}~\bibnamefont {Matsuo}}, \bibinfo {author} {\bibfnamefont
	  {T.}~\bibnamefont {Sakai}}, \bibinfo {author} {\bibfnamefont
	  {N.}~\bibnamefont {Katayama}}, \bibinfo {author} {\bibfnamefont
	  {H.}~\bibnamefont {Sawa}}, \bibinfo {author} {\bibfnamefont {K.}~\bibnamefont
	  {Kindo}}, \bibinfo {author} {\bibfnamefont {C.}~\bibnamefont {Michioka}},\
	  and\ \bibinfo {author} {\bibfnamefont {K.}~\bibnamefont {Yoshimura}},\ }\href
	  {https://doi.org/10.1103/PhysRevB.100.174401} {\bibfield  {journal} {\bibinfo
	   {journal} {Phys. Rev. B}\ }\textbf {\bibinfo {volume} {100}},\ \bibinfo
	  {pages} {174401} (\bibinfo {year} {2019})}\BibitemShut {NoStop}%
	\bibitem [{\citenamefont {Chen}\ \emph {et~al.}(2020)\citenamefont {Chen},
	  \citenamefont {Huang}, \citenamefont {Pan},\ and\ \citenamefont
	  {Mi}}]{chen2020quantum}%
	  \BibitemOpen
	  \bibfield  {author} {\bibinfo {author} {\bibfnamefont {X.-H.}\ \bibnamefont
	  {Chen}}, \bibinfo {author} {\bibfnamefont {Y.-X.}\ \bibnamefont {Huang}},
	  \bibinfo {author} {\bibfnamefont {Y.}~\bibnamefont {Pan}},\ and\ \bibinfo
	  {author} {\bibfnamefont {J.-X.}\ \bibnamefont {Mi}},\ }\href
	  {https://www.sciencedirect.com/science/article/abs/pii/S0304885319337953}
	  {\bibfield  {journal} {\bibinfo  {journal} {J. Magn. Magn. Mater.}\ }\textbf
	  {\bibinfo {volume} {512}},\ \bibinfo {pages} {167066} (\bibinfo {year}
	  {2020})}\BibitemShut {NoStop}%
	\bibitem [{\citenamefont {Liu}\ \emph {et~al.}(2022)\citenamefont {Liu},
	  \citenamefont {Yuan}, \citenamefont {Li}, \citenamefont {Li}, \citenamefont
	  {Zhao}, \citenamefont {Liao},\ and\ \citenamefont
	  {Li}}]{PhysRevB.105.024418}%
	  \BibitemOpen
	  \bibfield  {author} {\bibinfo {author} {\bibfnamefont {J.}~\bibnamefont
	  {Liu}}, \bibinfo {author} {\bibfnamefont {L.}~\bibnamefont {Yuan}}, \bibinfo
	  {author} {\bibfnamefont {X.}~\bibnamefont {Li}}, \bibinfo {author}
	  {\bibfnamefont {B.}~\bibnamefont {Li}}, \bibinfo {author} {\bibfnamefont
	  {K.}~\bibnamefont {Zhao}}, \bibinfo {author} {\bibfnamefont {H.}~\bibnamefont
	  {Liao}},\ and\ \bibinfo {author} {\bibfnamefont {Y.}~\bibnamefont {Li}},\
	  }\href {https://doi.org/10.1103/PhysRevB.105.024418} {\bibfield  {journal}
	  {\bibinfo  {journal} {Phys. Rev. B}\ }\textbf {\bibinfo {volume} {105}},\
	  \bibinfo {pages} {024418} (\bibinfo {year} {2022})}\BibitemShut {NoStop}%
	\bibitem [{\citenamefont {Zeng}\ \emph {et~al.}(2022)\citenamefont {Zeng},
	  \citenamefont {Ma}, \citenamefont {Wu}, \citenamefont {Li}, \citenamefont
	  {Tao}, \citenamefont {Lu}, \citenamefont {Chen}, \citenamefont {Mi},
	  \citenamefont {Song}, \citenamefont {Cao}, \citenamefont {Che}, \citenamefont
	  {Li}, \citenamefont {Li}, \citenamefont {Luo}, \citenamefont {Meng},\ and\
	  \citenamefont {Li}}]{PhysRevB.105.L121109}%
	  \BibitemOpen
	  \bibfield  {author} {\bibinfo {author} {\bibfnamefont {Z.}~\bibnamefont
	  {Zeng}}, \bibinfo {author} {\bibfnamefont {X.}~\bibnamefont {Ma}}, \bibinfo
	  {author} {\bibfnamefont {S.}~\bibnamefont {Wu}}, \bibinfo {author}
	  {\bibfnamefont {H.-F.}\ \bibnamefont {Li}}, \bibinfo {author} {\bibfnamefont
	  {Z.}~\bibnamefont {Tao}}, \bibinfo {author} {\bibfnamefont {X.}~\bibnamefont
	  {Lu}}, \bibinfo {author} {\bibfnamefont {X.-h.}\ \bibnamefont {Chen}},
	  \bibinfo {author} {\bibfnamefont {J.-X.}\ \bibnamefont {Mi}}, \bibinfo
	  {author} {\bibfnamefont {S.-J.}\ \bibnamefont {Song}}, \bibinfo {author}
	  {\bibfnamefont {G.-H.}\ \bibnamefont {Cao}}, \bibinfo {author} {\bibfnamefont
	  {G.}~\bibnamefont {Che}}, \bibinfo {author} {\bibfnamefont {K.}~\bibnamefont
	  {Li}}, \bibinfo {author} {\bibfnamefont {G.}~\bibnamefont {Li}}, \bibinfo
	  {author} {\bibfnamefont {H.}~\bibnamefont {Luo}}, \bibinfo {author}
	  {\bibfnamefont {Z.~Y.}\ \bibnamefont {Meng}},\ and\ \bibinfo {author}
	  {\bibfnamefont {S.}~\bibnamefont {Li}},\ }\href
	  {https://doi.org/10.1103/PhysRevB.105.L121109} {\bibfield  {journal}
	  {\bibinfo  {journal} {Phys. Rev. B}\ }\textbf {\bibinfo {volume} {105}},\
	  \bibinfo {pages} {L121109} (\bibinfo {year} {2022})}\BibitemShut {NoStop}%
	\bibitem [{\citenamefont {Hong}\ \emph {et~al.}(2022)\citenamefont {Hong},
	  \citenamefont {Behnami}, \citenamefont {Yuan}, \citenamefont {Li},
	  \citenamefont {Brenig}, \citenamefont {B\"uchner}, \citenamefont {Li},\ and\
	  \citenamefont {Hess}}]{PhysRevB.106.L220406}%
	  \BibitemOpen
	  \bibfield  {author} {\bibinfo {author} {\bibfnamefont {X.}~\bibnamefont
	  {Hong}}, \bibinfo {author} {\bibfnamefont {M.}~\bibnamefont {Behnami}},
	  \bibinfo {author} {\bibfnamefont {L.}~\bibnamefont {Yuan}}, \bibinfo {author}
	  {\bibfnamefont {B.}~\bibnamefont {Li}}, \bibinfo {author} {\bibfnamefont
	  {W.}~\bibnamefont {Brenig}}, \bibinfo {author} {\bibfnamefont
	  {B.}~\bibnamefont {B\"uchner}}, \bibinfo {author} {\bibfnamefont
	  {Y.}~\bibnamefont {Li}},\ and\ \bibinfo {author} {\bibfnamefont
	  {C.}~\bibnamefont {Hess}},\ }\href
	  {https://doi.org/10.1103/PhysRevB.106.L220406} {\bibfield  {journal}
	  {\bibinfo  {journal} {Phys. Rev. B}\ }\textbf {\bibinfo {volume} {106}},\
	  \bibinfo {pages} {L220406} (\bibinfo {year} {2022})}\BibitemShut {NoStop}%
	\bibitem [{\citenamefont {Lu}\ \emph {et~al.}(2022)\citenamefont {Lu},
	  \citenamefont {Yuan}, \citenamefont {Zhang}, \citenamefont {Li},
	  \citenamefont {Luo},\ and\ \citenamefont {Li}}]{lu2022observation}%
	  \BibitemOpen
	  \bibfield  {author} {\bibinfo {author} {\bibfnamefont {F.}~\bibnamefont
	  {Lu}}, \bibinfo {author} {\bibfnamefont {L.}~\bibnamefont {Yuan}}, \bibinfo
	  {author} {\bibfnamefont {J.}~\bibnamefont {Zhang}}, \bibinfo {author}
	  {\bibfnamefont {B.}~\bibnamefont {Li}}, \bibinfo {author} {\bibfnamefont
	  {Y.}~\bibnamefont {Luo}},\ and\ \bibinfo {author} {\bibfnamefont
	  {Y.}~\bibnamefont {Li}},\ }\href
	  {https://www.nature.com/articles/s42005-022-01053-4} {\bibfield  {journal}
	  {\bibinfo  {journal} {Commun. Phys.}\ }\textbf {\bibinfo {volume} {5}},\
	  \bibinfo {pages} {272} (\bibinfo {year} {2022})}\BibitemShut {NoStop}%
	\bibitem [{\citenamefont {Schulenburg}\ \emph {et~al.}(2002)\citenamefont
	  {Schulenburg}, \citenamefont {Honecker}, \citenamefont {Schnack},
	  \citenamefont {Richter},\ and\ \citenamefont
	  {Schmidt}}]{PhysRevLett.88.167207}%
	  \BibitemOpen
	  \bibfield  {author} {\bibinfo {author} {\bibfnamefont {J.}~\bibnamefont
	  {Schulenburg}}, \bibinfo {author} {\bibfnamefont {A.}~\bibnamefont
	  {Honecker}}, \bibinfo {author} {\bibfnamefont {J.}~\bibnamefont {Schnack}},
	  \bibinfo {author} {\bibfnamefont {J.}~\bibnamefont {Richter}},\ and\ \bibinfo
	  {author} {\bibfnamefont {H.-J.}\ \bibnamefont {Schmidt}},\ }\href
	  {https://doi.org/10.1103/PhysRevLett.88.167207} {\bibfield  {journal}
	  {\bibinfo  {journal} {Phys. Rev. Lett.}\ }\textbf {\bibinfo {volume} {88}},\
	  \bibinfo {pages} {167207} (\bibinfo {year} {2002})}\BibitemShut {NoStop}%
	\bibitem [{\citenamefont {Schnack}\ \emph {et~al.}(2020)\citenamefont
	  {Schnack}, \citenamefont {Schulenburg}, \citenamefont {Honecker},\ and\
	  \citenamefont {Richter}}]{PhysRevLett.125.117207}%
	  \BibitemOpen
	  \bibfield  {author} {\bibinfo {author} {\bibfnamefont {J.}~\bibnamefont
	  {Schnack}}, \bibinfo {author} {\bibfnamefont {J.}~\bibnamefont
	  {Schulenburg}}, \bibinfo {author} {\bibfnamefont {A.}~\bibnamefont
	  {Honecker}},\ and\ \bibinfo {author} {\bibfnamefont {J.}~\bibnamefont
	  {Richter}},\ }\href {https://doi.org/10.1103/PhysRevLett.125.117207}
	  {\bibfield  {journal} {\bibinfo  {journal} {Phys. Rev. Lett.}\ }\textbf
	  {\bibinfo {volume} {125}},\ \bibinfo {pages} {117207} (\bibinfo {year}
	  {2020})}\BibitemShut {NoStop}%
	\end{thebibliography}
\end{document}